# Spider Solitaire is NP-Complete

By Jesse Stern

**SpiderSolitaire**

This paper will prove that a generalized version of the popular solitaire variant SpiderSolitaire is NP-Complete. This proof is very similar to section 4.3 of Malte Helmert's Complexity results for standard benchmark domains in planning *Artificial Intelligence* 143 (2), pp. 219-262. 2003, but is re-tailored to fit the game of SpiderSolitaire, rather the FreeCell. In order to determine the domain of the game the rules of the original game must be understood. The rules can be summarized as follows:

SpiderSolitaire: The game is played with two full decks which results in 104 cards and 2 copies of the 4 suits. Each card begins in one of 10 *tableau piles*, 4 piles of 5 cards and 6 piles of 4 cards, or in the *deck*. The game is then played according to the following rules:
- Cards can be removed from the game if and only if they are in a pile with at least one card of each value in a certain suit, if they are all touching without any cards in between them, and if they are organized in numerical order with the lowest card of the suit on top and the highest valued card ending the set. If this case occurs all of the cards in the set must be removed from the game.
- Only the top card of any pile is face up, a new card can not be revealed until the card currently selected finds a new, acceptable pile to be placed upon.
- Cards may be picked up if they are on the top of a tableau pile or if they are the last card (when starting at the top of a tableau pile and moving down in the pile from there) in a set of same suit cards when every card after the t=*top*, when top is defined as top card of a tableau pile, card has a value of t + b where b is the number of cards on top of it. Only one card or sequence may be picked up at once.
- Cards and sets may only be placed on the top of a tableau pile if the pile is empty, or if the picked up card or the last card of the picked up sequences has a value of t – 1.
- If there are cards left in the deck, one card may be placed on the top of each tableau pile from the top of the deck, going from left to right.
- The player wins when all of the cards are removed from the game.

The average game of SpiderSolitaire uses a fixed deck size which only allows for a finite number of initial setups. This means a game of standard SpiderSolitaire can be decided in constant time given a computer that knows all of the initial set ups. To make the problem more interesting the deck size needs to be allowed to vary either by changing the number of suits or changing the number of cards in the existing suits. We must also assume that all cards and their order be visible to the player, as it would be impossible to tell the complexity otherwise.

Out of these given changing the suit length will be the best course of action. Another previously finite variable that will be expanded are tableau width and deck size, though deck size can also be a fixed constant.

**SpiderSolitaire task**

For a given integer $n \in \mathbb{N}$, $C_n = 2(\{\spadesuit, \clubsuit, \heartsuit, \diamondsuit,\}) \times \{1, \ldots, n\}$ is called the n-**deck**. Its elements are called **cards**. For a card (s, v), s is called its **suits** and v is called its **value**. A card is black if its suit is $\spadesuit$ or $\clubsuit$ and **red** if its suit is $\heartsuit$ or $\diamondsuit$.

An n-w-**tableau** is a set of at most w non-empty sequences over $C_n$ such that each card appears in at most two such sequences. The individual sequences are called **tableau piles**. The last card of the sequence is called **top** card, the subsequence that is obtained by removing the top card is called the **buried part** of the pile. A card c **matches** a tableau pile if and only if card c has a value of one less than the value of the top card.

A SpiderSolitaire is a 4-tuple (n, w, d, T), where
- $n \in \mathbb{N}$ is called the **suit length**,
- $w \in \mathbb{N}$ is called the **tableau width**,
- $d \in \mathbb{N}$ is called the **deck height**, and
- T is an n-w-tableau such that each card in $C_n$ appears in exactly one tableau pile. It is called the **initial tableau**.

These definitions need not be explained to anyone who understands SpiderSolitaire. An important issue to note is that tableau piles may not be of equal size, which is acceptable due to the fact that the tableau piles can be made the typical size simply by adding cards that can be immediately moved to the tops of these piles. Now the domain of legal SpiderSolitaire can be written.

**SpiderSolitaire Domain**

This domain matches SpiderSolitaire problems of suit length n and tableau width w to plans as follows.

The state T consists of all pairs and is an n-w-tableau. T is called the current tableau and it is assumed that any card not in T has been removed from the game. In the initial state, the current tableau is the initial tableau and the goal state is to have no cards in the current tableau.

There are two activities, pickup and drop. Both are partial functions on the state set that are not themselves actions of the planning task state model but helpful for defining them.

There are two kinds of pickup activities for a card c and two kinds for a sequence s. The first, pickup card from empty tableau, is defined in all states (T) in which c is the only card of some tableau pile p ϵ T and maps to (T \ {p}). The second, pickup card from non-empty tableau, is defined in all states (T) containing some tableau pile p with top card and non-empty buried part p' and maps to (T \ {p} U {p'}). The third pickup, pickup sequence from otherwise empty tableau, is defined in all states (T) in which s is the bottom card in a tableau pile which has only cards of the same suit which are decreasing in value by one when counting from s to the top card of the tableau pile (T \ {s}). The fourth pickup, pickup sequence from otherwise non-empty tableau, is defined in all states (T) in which s is the bottom card of a sequence in an otherwise non-empty tableau pile s' which has only cards of the same suit which are decreasing in value by one when counting from s to the top card of the tableau pile (T \ {s} U {s'}).

Additionally, there are two kinds of drop activities for card c and two kinds of drop activities for sequence s. Drop card on empty tableau and drop sequence on otherwise empty tableau are the inverse functions of pickup card from empty tableau and pickup sequence from otherwise empty tableau respectively. Drop card on non-empty tableau pile is defined in all states (T) where the card currently picked up u has a value one less than the top card c of some tableau pile p and maps to (T \ {u} U {c}). Drop sequence on non-empty tableau is defined in all states (T) where the bottom card of the sequence u has a value one less than the top card c of some tableau pile p and maps to (T \ {u} U {c}).

Now that pickup and drop activities have been formally defined, the set of actions of the planning task state models consists of all compositions drop°pickup.

**PROOF: SpiderSolitaire is in NP**

The only kinds of actions that cannot be immediately reversed by applying an inverse action are movements from tableau pile where the top card does not match the buried part of the pile and movements that create full sets (causing the cards to be removed from the board).

As cards that do not match the card they are on top of is an instance that only occurs in the set up of the the initial tableau, or by cards being played from the deck if there is a deck, so the number of actions of the first kind is bounded by the number of these in the initial tableau plus the number of cards in the deck (as it is possible that any number of these cards may be placed on top of a card it does not match when dealt). Cards that are out of play once they have formed a complete set, so the number of actions of the second kind in any plan is bounded by the total number of cards.

Because there is a polynomial bound on the number of non-undoable movements the number of actions in shortest plans are polynomial bounded.

It can also be said that the only types of movements that bring the SpiderSolitaire problem closer to completion are moves that place cards in to sets that will later clear and, because cards required to complete these sets are often buried, the movement any number of cards to uncover a card that will then be moved to a set. Because there are only 8n cards there are only 8n-8 moves that will result in a card being moved to a set as the highest cards in a suit need not move. Any time a card is moved to a set the only hindrance are cards on top of it which must be moved out of the way first. If these cards can be moved off of the card attempting to be moved to a set, then they can be done so in at most 8n-1, as you will only ever have to move a card once for it to be out of the way of the card being moved to a set. It follows that if it takes 8n-1 moves, at most, to make any of the 8n-8 moves of cards to there set, if no cards start out in a partially completed set, then it follows that any solution to the problem can not take more then $64n^2-72n+8$ moves. This bounds any given solution to a SpiderSolitaire problem in polynomial time.

**Proof: SpiderSolitaire is in NP-Hard**

This proof is by reduction from 3SAT. Let (V, C) be a 3SAT instance, where $V = \{v_1, \ldots, v_n\}$ is a set of variables and $C = \{c_1, \ldots, c_n\}$ is a set of clauses over V which contain exactly three literals each. We write $l_{i,j}$ for the j-th literal of the i-th clause. The corresponding SpiderSolitaire task is rather intricate and the reader should view Figure 1 to better understand how the propositional formula and planning tasks are interrelated.

To order the literals over V, we designate $v_i$ as the (2i – 1)-th literal, $¬v_i$ as the 2i-th literal and $l_k$ as the k-th literal. We define the number of instances of $l_k$ as the number of pairs (I, j) such that $l_{i,j} = l_k$, and the corresponding pairs are called the {first, second, . . . , n} occurrence of $l_k$. Additionally, we define $o_k$, the cumulated number of occurrences up to $l_k$, as the sum of the number of occurrences of $l_{k'}$ for all k' ≤ k.

We now define the selection value $val_S$ as $|C| + 2n + 2$, the literal value of $l_k$ as $val_k = val_S + 2k + 2o_k$, the clause value $val_C$ as $val_{2n} + 2$ and the bottom value $val_B$ as $val_C + 6|C|$. In Figure 1, $val_S$ = 11, $val_1$ = 15 (for literal $v_1$), $val_2$ = 21 (for $v_1$), $val_3$ = 27 (for $v_2$), $val_4$ = 31 (for $v_2$), $val_5$ = 37 (for $v_3$), $val_6$ = 37 (val $v_3$), $val_C$ = 43, $val_B$ = 61 and $val_F$ = $val_B$ + 4|C|.

The SpiderSolitaire task has a suit length of $val_B + 4|C|$, a tableau width of $6|C| + 2|V| + 10$ and a deck count of 0. The initial tableau is shown in Figure 1 below.

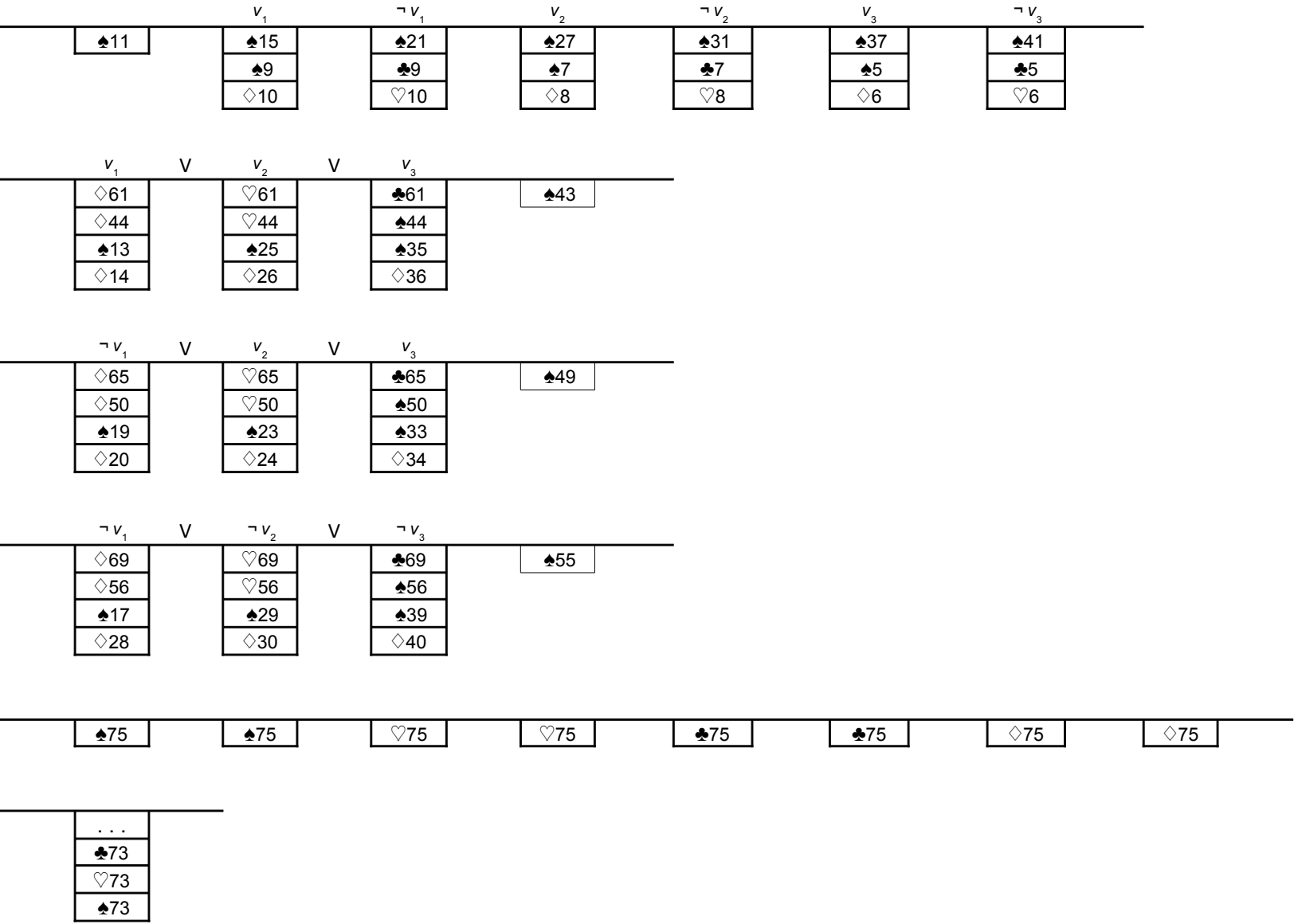

The piles of the initial tableau, depicted in Figure 1, fall into on of four groups.  The first 2|V| + 1 piles are called the literal selection piles, depicted at the top of the figure.  The next 4|C| piles are called the clause piles, organized into subgroups of six piles that each relate to a specific clause, called clause groups.  The next 8 piles are called the foundation piles, shown as the second to last group of cards in the figure.  The last pile, holding most of the cards and all of the duplicates (excluding all card with value val_F), is called the big pile.  Note that cards at the top of a pile are shown near the bottom of the picture, following the usual convention of SpiderSolitaire programs on computers.

**Literal selection piles:**  The first literal selection pile contains only the card (♠, val_s).  The other piles contain three cards each, Each of which corresponds to a different literal.  The pile for $l_k$ is defined as (♠, val_k) (♠, val_s − k − 1)(♢, val_s − k) if k is odd and as (♠, val_k) (♣, val_s − k)( ♡, val_s − k + 1) if k is even.

**Clause groups:** Each group is organized as follows. There are six piles corresponding to the i-th clause. We set the *bottom* value for the group as bottom = $val_B + 4(i - 3)$ and the base value for the group as base = $val_C + 6(i - 3)$.

The first three piles contain four cards each. The first and second of these are of value bottom and base + 1, respectively, suit does not matter. The remaining cards are dependent on the literals in the clause: For $1 \leq j \leq 3$, the third and fourth cards of the j-th pile are ($\spadesuit$, $val_k - 2m$) and ($\diamondsuit$, $val_k - 2m$), where k and m are calculated such that (i, j) is the m-th occurrence of $l_k$. The fourth pile contains only one card and is defined as.

**Foundation piles:** The eight foundation piles all have one card in them and the cards are as follows from left to right, ($\spadesuit$, $val_F$) ($\spadesuit$, $val_F$) ($\heartsuit$, $val_F$) ($\heartsuit$, $val_F$) ($\clubsuit$, $val_F$) ($\clubsuit$, $val_F$) ($\diamondsuit$, $val_F$) ($\diamondsuit$, $val_F$).

**Big pile:** The top i cards of the this pile are any card, that has not appeared twice, that cannot be moved by any means other than being moved to the sixth pile of a clause once that pile has been emptied of cards. The easiest way to assure these cards are not moved early is to assign them a value of $val_B + 4|C| - 2$.

We now show that this SpiderSolitaire task can be solved if and only if there is a satisfying assignment to the variables of the logical formula. First, assume there is such a satisfying assignment a: $V \rightarrow \{ \top, \bot \}$. The following strategy solves the task:

For each i $\in$ {1, . . . , n}, move the top two cards from the literal selection piles that correspond to literals which are true under a to the first literal selection pile. This releases the bottom cards of some literal selection piles, spades cards which can then be used to move cards from the clause piles. In the example of Figure 1, for the assignment {($v_1$, $\bot$), ($v_1$, $\bot$), ($v_1$, $\top$)} these are the 15, 27 and 41 of spades. These are called the literal choice cards.

The first three piles of each clause group correspond to the literals in that clause. The top two cards of such a pile can be moved to the literal selection piles if and only if the literal choice card of the corresponding literal has been revealed.

Because the task given has a satisfying truth assignment, a literal is satisfied in each clause, and thus it is possible to remove the top two cards of one of the first three piles of each clause group. If any literal is satisfied in a clause it allows the card in the fourth pile to be moved.

After this has been done for all clauses, the first |C| cards can be moved to empty tableaus. This reveals many high- valued cards. The problem then solves itself as all of the cards begin to move in to numeric order on top of the foundation piles.

Now assume that the SpiderSolitaire task is solvable. It is not possible to move the bottom card of any tableau pile within the tableau, other than the fourth tableau in a clause, before the top card of the big pile is moved, because all cards that they could be moved on top of are buried in the big pile. This implies that the first i movements of the top card of the big pile cannot go to an empty tableau position other than the fourth tableau in a clause.

On the other hand, it can not be moved on top of any other card as its first movement, because all possible destination cards are buried under it. Together, this implies that its first (and thus only) movement must be directly to one of the empty forth piles in a clause.

For each clause group, the top card of the fourth pile must be moved, and it can only be moved to the third card of the first, second, or third clause pile, requiring the top card of one of these piles to be moved. Thus, in each clause group, the top two cards of one of the first three piles must be moved somewhere else for the task to be solvable.

The only way this can be done is by uncovering the literal choice cards of corresponding literals in the way explained in the other direction of the proof. As it is not possible to uncover the literal choice card for $v_1$ and $\neg v_1$ at the same time (for any i), this requires the existence of a satisfying assignment to the truth variables, completing the proof.

If the deck height is larger than zero, a very similar reduction can be used by ensuring that all cards from the deck must be moved to foundations and that the suit length is increased accordingly. This concludes the proof NP-Hardeness. Now that SpiderSolitaire has been proven to be in both NP and NP-Hard, we know it to be in NP-Complete.

**Refrences**
[1]     **Complexity results for standard benchmark domains in planning**.
        *Artificial Intelligence* 143 (2), pp. 219-262. 2003.